\documentclass[10pt,letterpaper]{article}
\usepackage{amsmath}
\usepackage{amssymb}
\usepackage{cite}

\usepackage{opex3}
\usepackage{txfonts}
\usepackage{amssymb}
\usepackage{tipa}
\usepackage{graphicx}
\usepackage{txfonts}
\usepackage{amssymb}

\begin{document}

\title{Tunable multi-channel inverse optomechanically induced transparency
and its applications}
\author{Qin Wu,$^{1,2,3}$ Jian-Qi Zhang,$^{2,5}$ Jin-Hui Wu,$^4$ Mang Feng,$
^{2,6}$  and Zhi-Ming Zhang$^{1*}$}
\address{$^1$Guangdong Provincial Key Laboratory of Nanophotonic Functional Materials and Devices (SIPSE), and
Guangdong Provincial Key Laboratory of Quantum Engineering and Quantum Materials, South China
Normal University, Guangzhou 510006, China\\
$^2$State Key Laboratory of Magnetic Resonance and Atomic and Molecular Physics, Wuhan Institute of Physics and Mathematics,
Chinese Academy of Sciences, Wuhan 430071, China \\
$^3$School of Information Engineering, Guangdong Medical University, Dongguan 523808, China\\
$^4$Center of Quantum Science, Northeast Normal University, Changchun 130117, China\\
$^{5}$changjianqi@gmail.com\\
$^{6}$ mangfeng@wipm.ac.cn}

\email{$^{*}$ zmzhang@scnu.edu.cn}

\begin{abstract}
In contrast to the optomechanically induced transparency (OMIT) defined
conventionally, the inverse OMIT behaves as coherent absorption of the input
lights in the optomechanical systems. We characterize a feasible inverse
OMIT in a multi-channel fashion with a double-sided optomechanical cavity
system coupled to a nearby charged nanomechanical resonator via Coulomb
interaction, where two counter-propagating probe lights can be absorbed via
one of the channels or even via three channels simultaneously with the
assistance of a strong pump light. Under realistic conditions, we
demonstrate the experimental feasibility of our model by considering two
slightly different nanomechanical resonators and the possibility of
detecting the energy dissipation of the system. In particular, we find that
our model turns to be a unilateral inverse OMIT once the two probe lights
are different with a relative phase, and in this case the relative phase can
be detected precisely.
\end{abstract}






\section{Introduction}

Electromagnetically induced transparency (EIT) \cite{prl-64-1107} is caused
by quantum interference, creating a narrow transmission window within an
absorption line. EIT was first theoretically predicted in three-level atoms
\cite%
{pra-24-379,pra-34-4785,prl-47-408,josab-1-102,pra-35-3354,pra-35-4592,pr-190-I}
and then observed in optically opaque strontium vapor \cite%
{prl-66-2593,prl-67-3062}. So far, EIT effects have attracted considerable
attention both theoretically and experimentally due to relevant optical
effects and applications, such as, optical Kerr effect and optical switch
\cite{nature-432-482,prl-110-113902}, slow light and quantum memory \cite%
{pra-46-R29,prl-82-4611,nature-397-594}, and quantum interference and
vibrational cooling \cite{po-35-259,prl-77-4326}. The key point in
realization of the EIT effect is to find a $\Lambda $-type level
configuration and construct quantum interference. In this context, for some
hybrid systems with $\Lambda $-type level structures, e.g., metal materials
\cite{PRL-101-047401}, coupled waveguides \cite {josab-22-1062,pra-88-013813,nc-5-5082},
atom-cavity systems \cite {nature-465-755}, and optomechanical systems \cite
{pra-83-043826,JPB-44-165505,CPB-22-024204,pra-88-013804,OL-39-4180,
pra-86-013815,OE-23-11508,optica-1-425,srep-5-9663,pra-81-041803,pra-86-053806},
the analog of the EIT effects can also be observed.

The analog of the EIT effects in the optomechanics is named as the
optomechanically induced transparency (OMIT), which was predicted in a
pioneering theoretical work \cite{pra-81-041803}, and then verified
experimentally \cite{pra-88-013813,science-330-1520}. Very recently, not
only the slow light was experimentally confirmed in the OMIT system \cite
{nature-472-69}, but also the optomechanical dark mode was observed
experimentally \cite{science-338-1609}. Motivated by these experiments, many
different proposals \cite
{pra-86-053806,jpb-46-025501,pra-81-033830,nature-471-204,pra-87-013824,
pra-85-021801, JPB-47-055504, arxiv,NJP-16-033023,OE-22-4886, pra-90-043825}
were proposed based on the OMIT effect. One of the outstanding works is for
an inverse OMIT \cite{NJP-16-033023} in an optomechanical cavity, i.e., an
optomechanical resonator inside a single-mode cavity, which shows that, when
two weak counter-propagating probe lights within the narrow transmission
window of the OMIT are injected simultaneously, neither of the probe lights
can be output from the cavity due to complete absorption by the
optomechanics. Therefore, this effect is also named to be the coherent
perfect absorption and has been stretched to two optomechanical cavities
coupled to an optomechanical resonator \cite{OE-22-4886}, showing the
prospective for coherent perfect transmission and beyond. However, both the
schemes \cite{NJP-16-033023,OE-22-4886} are very hard to achieve
experimentally due to stringent conditions involved. As a result, it is
desirable to have an experimentally feasible scheme for demonstrating the
inverse OMIT. In addition, exploring the applications of the inverse OMIT is
also interesting and experimentally demanded.

On the other hand, with an optomechanical cavity coupled to an external
nanomechanical resonator (NR) via Coulomb interaction, the single narrow
transmission window in the output light is split into two narrower
transmission windows with the splitting governed by the Coulomb coupling
\cite{pra-90-043825}. This is due to the fact that an additional hybrid
energy level is introduced into the original three-level system by the
Coulomb coupling between the external NR and the optomechanical resonator.
Similar double OMIT effect can also be observed when the optomechanical
resonator interacts with a qubit \cite{pra-90-023817} or an NR \cite%
{pra-91-063827, nphotonics, apl-105-014108}. The Coulomb interaction works
for a wide range from nanometer to meter \cite{pra-90-043825,pra-72-041405,
pra-91-022326} and can be controlled by the bias voltage \cite%
{pra-90-043825,pra-91-022326}. Besides, it can also be applied to different
kinds of charged objects at different frequencies \cite{pra-72-041405}.
These advantages remind us of the necessity to explore a multi-channel
inverse OMIT in the optomechanical system with the tunable Coulomb
interaction.

In the present work, by considering a double-sided optomechanical cavity
(involving a charged NR) coupled to another identical charged NR nearby via
Coulomb interaction, we present a multi-channel inverse OMIT and study the
energy dissipation of the system through the intracavity photon number and
the mechanical excitations of the NRs. In addition, we explore a unilateral
inverse OMIT, i.e., observation of the inverse OMIT available only on one
side of the optomechanical cavity. Our study shows that this unilateral
inverse OMIT could be applied to precision measurement of the relative phase
between two probe lights.

Compared with previous studies, our idea includes more interesting physics
and thus owns different applications. First, our inverse OMIT is generated
from a double-OMIT system. It is a multi-channel inverse OMIT with the
windows of narrower profiles than the counterpart in \cite%
{NJP-16-033023,OE-22-4886}, and the dissipation of the input probe light can
be directly detected by the external NR, without the need of an additional
light field as required in \cite{NJP-16-033023,OE-22-4886}. Second, if there
is a relative phase between two probe lights, the inverse OMIT is observed
only on one side of the optomechanical cavity, which is essentially
different from the inverse OMIT in \cite{NJP-16-033023,OE-22-4886}. This
unilateral inverse OMIT can not only reduce the experimental difficulty for
demonstrating the inverse OMIT effects, but also be very sensitive to the
relative phase between the two probe lights. As such, it can be applied to
detect the relative phase between two probe lights. In addition, different
from the analog of electromagnetically induced absorptions realized by a
Stokes process in the region of blue detuning \cite%
{nature-472-69,NJP-14-123037}, our scheme can be achieved by an anti-Stokes
process in the region of red detuning. We argue below that these
characteristics might be helpful for practical applications using
optomechanical systems.

The present paper is structured as follows. Sec. 2 presents the model and an
analytical solution to the multi-channel inverse OMIT of an optomechanical
system. In Sec. 3, we characterize the output probe fields. Under some
realistic conditions, we explore in Sec. 4 the situation of non-identical
NRs, the energy dissipation and a possible application in precision
measurement. The experimental feasibility is discussed in Sec. 5. A
brief conclusion can be found in the last section.

\section{The model and solution}

\begin{figure}[tbp]
\centering
\includegraphics[width=8cm]{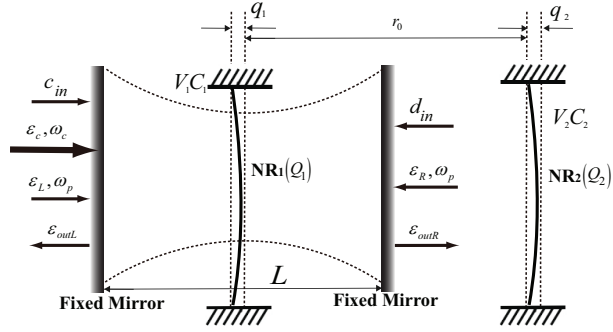}
\caption{Schematic diagram for a double-sided cavity with a nanomechanical
resonator NR$_{1}$ located at the node of the cavity mode and a
nanomechanical resonator NR$_{2}$ outside. NR$_{1}$ is charged by the bias
gate voltage $V_{1}$ and subject to the Coulomb force due to another charged
NR$_{2}$ with the bias gate voltage $V_{2}$. The optomechanical cavity of
length $L$ is driven by three light fields, one of which is the pump field $%
\protect\varepsilon _{c}$ with frequency $\protect\omega _{c}$ and the other
of which are the probe fields $\protect\varepsilon _{L(R)}$ with frequency $%
\protect\omega_{p}$. The output field is represented by $\protect\varepsilon %
_{outL(R)}$. $q_{1}$ and $q_{2}$ represent the small displacements of NR$%
_{1} $ and NR$_{2}$ from their equilibrium positions, and $r_{0}$ is the
equilibrium distance between the two charged NRs.}
\label{fig1}
\end{figure}

As sketched in Fig. \ref{fig1}, our system consists of a Fabry-Perot (FP)
cavity and two charged NRs, i.e., NR$_{1}$ and NR$_{2}$. The NR$_{1}$ is
inside the FP cavity formed by two fixed mirrors with finite equal
transmissions, and couples to the cavity mode with a radiation pressure. The
NR$_{1}$ also interacts with the NR$_{2}$ outside the FP cavity via a
tunable Coulomb interaction. We suppose that the FP cavity is driven by a
strong pump field (frequency $\omega _{c}$) from the left-hand side of the
cavity, and two weak classical probe fields (frequency $\omega _{p}$) are
injected into the cavity from both sides of the cavity. The Hamiltonian in
the rotating frame at the pump field frequency $\omega _{c}$ can be written
as \cite{pra-90-043825,pra-91-063827}
\begin{eqnarray}
H &=&\hbar (\omega _{0}-\omega _{c})c^{\dag }c+(\frac{p_{1}^{2}}{2m_{1}}+%
\frac{1}{2}m_{1}\omega _{1}^{2}q_{1}^{2})+(\frac{p_{2}^{2}}{2m_{2}}+\frac{1}{%
2}m_{2}\omega _{2}^{2}q_{2}^{2})+i\hbar \varepsilon _{c}(c^{\dag }-c)  \notag
\\
&+&i\hbar (c^{\dag }\varepsilon _{L}e^{-i\delta t}-h.c.)+i\hbar (c^{\dag
}\varepsilon _{R}e^{-i\delta t}-h.c.)+\hbar g_{0}c^{\dag }cq_{1}+\hbar
\lambda _{0}q_{1}q_{2},  \label{M0}
\end{eqnarray}
where the first three terms represent the free parts of the Hamiltonian for
the cavity field and the NRs. $c\ (c^{\dag })$ is the annihilation
(creation) operator of the cavity mode at frequency $\omega _{0}$. The
charged NR$_{1}$ (NR$_{2}$) owns the frequency $\omega _{1}$ ($\omega _{2}$%
), the effective mass $m_{1}$ ($m_{2}$), the position $q_{1}$ ($q_{2}$) and
the momentum $p_{1}$ ($p_{2}$) \cite{pra-85-021801}. The next three terms
describe the cavity mode driven by a pump field and two probe fields. $%
\varepsilon _{c}=\sqrt{2\kappa \wp _{c}/\hbar \omega _{c}}$ ($\varepsilon
_{L(R)}=\sqrt{2\kappa \wp _{p}/\hbar \omega _{p}}$) is an amplitude of the
strong pump (weak probe) field with $\wp _{c}$ ($\wp _{p}$) and $\kappa $
being the power of the pump (probe) field and the cavity decay rate,
respectively, and $\delta =\omega _{p}-\omega _{c}$ is a detuning between
the probe field and the pump field. The last two terms include the coupling
of the NR$_{1}$ to the cavity mode via the radiation pressure strength $%
g_{0} $ \cite{pra-77-033819}, and also the interaction between the NR$_{1}$
and NR$_{2}$ with the Coulomb coupling strength $\lambda _{0}=\frac{%
C_{1}V_{1}C_{2}V_{2}}{2\pi \hbar \varepsilon _{0}r_{0}^{3}}$\cite%
{pra-72-041405, arxiv:1402.6434}. The NR$_{1}$ (NR$_{2}$) takes the charge $%
Q_{1}=C_{1}V_{1}$ ($Q_{2}=-C_{2}V_{2}$), with $C_{1}(C_{2})$ and $%
V_{1}(-V_{2})$ being the capacitance and the voltage of the bias gate,
respectively.

With the annihilation (creation) operator $b_{j}$ ($b_{j}^{\dagger}$), the
position and momentum operators of the NR$_{j}$\ are rewritten as $q_{j}=
\sqrt{\frac{\hbar }{2m_{j}\omega _{j}}}(b_{j}+b_{j}^{\dagger})$, and $
p_{j}=i\sqrt{\frac{\hbar m_{j}\omega _{j}}{2}}(b_{j}^{\dagger}-b_{j})$, which yields the
Hamiltonian
\begin{eqnarray}
&&H^{\prime }=\hbar \Delta _{c}c^{\dag }c+\hbar \omega _{1}b_{1}^{\dagger
}b_{1}+\hbar \omega _{2}b_{2}^{\dagger }b_{2}  \notag  \label{MM0} \\
&&+\hbar gc^{\dag }c(b_{1}+b_{1}^{\dagger })+i\hbar \varepsilon _{c}(c^{\dag
}-c)+\hbar \lambda (b_{1}^{\dagger }+b_{1})(b_{2}^{\dagger }+b_{2})  \notag
\\
&&+i\hbar (c^{\dag }\varepsilon _{L}e^{-i\delta t}-h.c.)+i\hbar (c^{\dag
}\varepsilon _{R}e^{-i\delta t}-h.c.),
\end{eqnarray}%
with $\Delta _{c}=\omega _{0}-\omega _{c}$, $g=g_{0}\sqrt{\hbar
/2m_{1}\omega _{1}}$\ and $\lambda =\frac{\hbar \lambda _{0}}{2}\sqrt{%
m_{1}m_{2}\omega _{1}\omega _{2}}$.

Considering the damping and noise terms, the quantum Langevin equations are
generated from Eq. (\ref{MM0}),
\begin{equation}
\begin{array}{ccl}
\dot{b_{1}} & = & -(\frac{\gamma _{1}}{2}+i\omega _{1})b_{1}-igc^{\dag
}c-i\lambda (b_{2}^{\dagger }+b_{2})+\sqrt{\gamma _{1}}b_{1}^{in}, \\
\dot{b_{2}} & = & -(\frac{\gamma _{2}}{2}+i\omega _{2})b_{2}-i\lambda
(b_{1}^{\dagger }+b_{1})+\sqrt{\gamma _{2}}b_{2}^{in}, \\
\dot{c} & = & -(2\kappa +i\Delta _{c})c-ig(b_{1}+b_{1}^{\dag })c+\varepsilon
_{c}+(\varepsilon _{L}+\varepsilon _{R})e^{-i\delta t}+\sqrt{2\kappa }%
(c_{in}+d_{in}),%
\end{array}
\label{MM01}
\end{equation}%
where $\gamma _{1}$\ ($\gamma _{2}$) is the NR$_{1}$\ (NR$_{2}$) decay rate,
$2\kappa $\ is the cavity decay rate from the two sides. The quantum
Brownian noise $b_{1}^{in}$\ $(b_{2}^{in})$\ is resulted from the coupling
between the NR$_{1}$\ (NR$_{2}$) and the environment \cite{pra-77-033804}. $%
c_{in}(d_{in})$\ is the input quantum noise from the environment \cite%
{pra-77-033804}. The mean values of the noise terms $b_{1}^{in}$, $b_{2}^{in}
$, $c_{in}$, and $d_{in}$\ are zero.

Equation (\ref{MM01}) is solved under the conditions: (i) The sideband resolved regime ($\omega
_{1}\gg \kappa $); (ii) $\delta \simeq \omega _{1}$\ and $\delta \simeq
\omega _{2}$; (iii) $\Delta \sim \omega _{1}$. The first condition ensures
an observable normal mode splitting due to strong coupling
between the NR$_{1}$ and the cavity mode; The second condition yields
$\delta ^{2}-\omega _{1}^{2}=2\omega _{1}(\delta -\omega
_{1})$\ and $\delta ^{2}-\omega _{2}^{2}=2\omega _{2}(\delta -\omega _{1})$;
The last one is to eliminate the detuning $\Delta$.
We also suppose that each operator is a mean value plus a small
quantum fluctuation, i.e., $o=o_{s}+\delta o$, with $o=b_{1}$, $b_{2}$, and $%
c$, and $\delta o\ll |o_{s}|$. Inserting them into Eq. (\ref{MM01}) and
neglecting the second-order smaller terms, we obtain the steady-state mean
values of the system as\emph{\ }%
\begin{equation}
\begin{array}{lll}
b_{1s} & =\frac{-ig|c_{s}|^{2}}{\frac{\gamma _{1}}{2}+i\omega _{1}+\frac{%
8\lambda ^{2}\omega _{1}\omega _{2}}{(\omega _{1}+i\frac{\gamma _{1}}{2}%
)(\omega _{2}+i\frac{\gamma _{2}}{2})(\frac{\gamma _{2}}{2}+i\omega _{2})}}
& \simeq \frac{-ig|c_{s}|^{2}}{\frac{\gamma _{1}}{2}+i\omega _{1}+\frac{%
8\lambda ^{2}}{\frac{\gamma _{2}}{2}+i\omega _{2}}}, \\
b_{2s} & =\frac{-2\lambda \omega _{1}}{(\omega _{1}+i\frac{\gamma _{1}}{2}%
)(\omega _{2}-i\frac{\gamma _{2}}{2})}\simeq \frac{-i2\lambda }{\frac{\gamma
_{2}}{2}+i\omega _{2}}b_{1s}, & c_{s}=\frac{\varepsilon _{c}}{2\kappa
+i\Delta }%
\end{array}%
\end{equation}%
with $\Delta =\omega _{0}-\omega _{c}+g(b_{1s}+b_{1s}^{\ast })$, and the
corresponding linearized quantum Langevin equations for the small quantum
fluctuations are of the form\emph{\ }%
\begin{equation}
\begin{array}{ccl}
\dot{\delta b}_{1} & = & -({\frac{{\gamma _{1}}}{2}+}i{\omega _{1}})\delta {%
b_{1}}-i({G^{\ast }\delta c+G\delta c^{\dagger })}-i\lambda (\delta {b_{2}+}%
\delta b_{2}^{\dagger })+\sqrt{{\gamma _{1}}}b_{1}^{in}, \\
\dot{\delta b}_{2} & = & -(\frac{{\gamma _{2}}}{2}+i{\omega _{2}})\delta {%
b_{2}}-i\lambda (\delta {b_{1}+}\delta b_{1}^{\dagger })+\sqrt{{\gamma _{2}}}%
b_{2}^{in}, \\
\dot{\delta c} & = & -(2\kappa +i\Delta )\delta c-iG(\delta {b_{1}}+\delta
b_{1}^{\dagger })+({\varepsilon _{L}}+{\varepsilon _{R}}){e^{-i\delta t}}+%
\sqrt{2\kappa }({c_{in}}+{d_{in}}),%
\end{array}
\label{bb}
\end{equation}%
with $G=gc_{s}$\ being the effective radiation pressure coupling.

The inverse OMIT effect can be studied by analyzing the mean response of the
system to two probe fields in the presence of the pump field. After input
noises of the system are ignored, the mean value equations with the probe
fields in Eq. (\ref{bb}) are rewritten as \cite{pra-81-041803,pra-86-053806}%
\emph{\ }%
\begin{equation}
\begin{array}{ccl}
\langle \dot{\delta b}_{1}\rangle  & = & -i({G^{\ast }}\left\langle {\delta c%
}\right\rangle {+G}\left\langle {\delta c^{\dagger }}\right\rangle {)}-(i{%
\omega _{1}+\frac{{\gamma _{1}}}{2}})\langle \delta {b_{1}}\rangle -i\lambda
(\langle \delta {b_{2}}\rangle +\langle \delta b_{2}^{\dagger }\rangle ), \\
\langle \dot{\delta b}_{2}\rangle  & = & -(i{\omega _{2}}+\frac{{\gamma _{2}}%
}{2})\langle \delta {b_{2}}\rangle -i\lambda (\langle \delta {b_{1}}\rangle
+\langle \delta b_{1}^{\dagger }\rangle ), \\
\langle \dot{\delta c}\rangle  & = & -(2\kappa +i\Delta )\langle \delta
c\rangle -iG(\langle \delta {b_{1}}\rangle +\langle \delta b_{1}^{\dagger
}\rangle )+({\varepsilon _{L}}+{\varepsilon _{R}}){e^{-i\delta t}}.%
\end{array}
\label{L1}
\end{equation}

Define the solution to Eq. (\ref{L1}) takes the following form \cite%
{pra-86-053806}
\begin{equation}
\langle \delta o\rangle =\delta o_{+}e^{-i\delta t}+\delta {o}_{-}e^{i\delta
t},  \label{MM1}
\end{equation}
the results for the small quantum fluctuations is given by
\begin{equation}
\begin{array}{ccl}
\delta b_{1+} & = & \frac{-iG^{\ast }}{\frac{{\gamma _{1}}}{2}-i(\delta -{%
\omega _{1}})+\frac{{\lambda ^{2}}}{{\frac{{\gamma _{2}}}{2}}-i(\delta -{%
\omega _{2}})}}\delta c_{+}, \\
\delta b_{2+} & = & \frac{-i\lambda }{\frac{\gamma _{2}}{2}+i\omega -i\delta
}\delta b_{1+}, \\
\delta c_{+} & = & \frac{\varepsilon _{L}+\varepsilon _{R}}{2\kappa
+i(\Delta -\delta )+\frac{|G|^{2}}{A}},
\end{array}
\label{M2}
\end{equation}
with $A=\frac{{\gamma _{1}}}{2}-i(\delta -\omega _{1})+\frac{{\lambda ^{2}}}{%
{\frac{{\gamma _{2}}}{2}}-i(\delta -{\omega _{2}})}$. Our scheme includes a
more general situation than in \cite{NJP-16-033023} since our results can be
reduced to the counterpart in \cite{NJP-16-033023} in the absence of the
Coulomb coupling. It is confirmed in the comparison with the output
field involving two tunable central frequencies for the inverse OMIT in \cite%
{NJP-16-033023} that our scheme owns three frequencies for the inverse OMIT
effect, two of which can be adjusted by the Coulomb interaction.

\section{The inverse OMIT}

Based on the solutions above, we present below the multi-channel inverse
OMIT in our system with some channels controllable by the driven field due
to effective coupling and Coulomb interaction between the NRs.

For simplicity, we first assume two identical charged NRs ($\omega
_{1}=\omega _{2}=\omega _{m}$) and the detuning between the pump field and
the cavity mode satisfying $\Delta\simeq\omega_{m}$. This assumption helps
an analytical understanding of characteristics of the model, but changes
nothing in the physical essence of the considered system. The assumption
will be released later under consideration of realistic condition.

Considering the output fields from the two sides of the cavity by the
input-output relations \cite{quantum}
\begin{equation}
\varepsilon _{out\alpha }+\varepsilon _{\alpha }e^{-iDt}=2\kappa \langle
\delta c\rangle ,~~~\alpha =R,L,  \label{M3}
\end{equation}
with $D=\delta -\omega _{m}$, we define the output fields as
\begin{equation}
\varepsilon _{out\alpha }=\varepsilon _{out\alpha +}e^{-iDt}+\varepsilon
_{out\alpha -}e^{iDt},~~~\alpha =R,L,  \label{M4}
\end{equation}
where $\varepsilon _{out\alpha +}$ and $\varepsilon _{out\alpha -}$ are the
responses at the frequencies $\omega _{p}$ and $2\omega _{c}-\omega _{p}$ in
the original frame.

Using Eqs. (\ref{MM1}), (\ref{M3}) and (\ref{M4}), the output fields at the
probe frequency $\omega _{p}$ are presented as
\begin{equation}
\begin{array}{rl}
\varepsilon _{out\alpha +}= & 2\kappa \delta {c}_{+}-\varepsilon _{\alpha }
\\
= & \frac{\varepsilon _{L}+\varepsilon _{R}}{2\kappa +iD+\frac{|G|^{2}}{
\frac{{\gamma _{1}}}{2}-i(\delta -\omega _{1})+\frac{{\lambda ^{2}}}{{\frac{{%
\ \gamma _{2}}}{2}}-iD}}}-\varepsilon _{\alpha}%
\end{array}
~~~  \label{M44}
\end{equation}
with $\alpha =R,L$.

\begin{figure}[tbp]
\centering
\includegraphics[width=8cm]{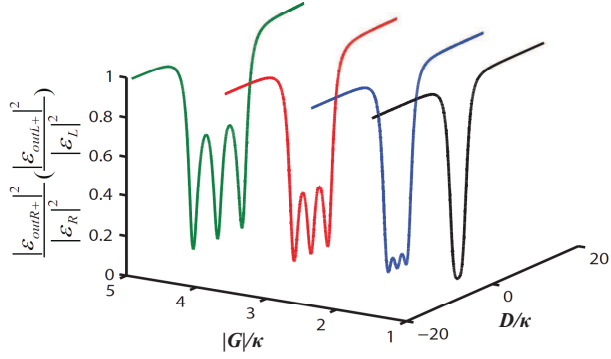}
\caption{The normalized output probe photon number $|\frac{
\protect\varepsilon _{outR+}}{\protect\varepsilon _{L}}|^{2}$ $(|\frac{
\protect\varepsilon _{outL+}}{\protect\varepsilon _{L}}|^{2})$ as functions
of the probe detuning $D/\protect\kappa $ and the effective radiation
coupling $|G|/\protect\kappa $, where $D=\protect\omega _{p}-\protect\omega %
_{c}-\protect\omega_{m}$. }
\label{fig2}
\end{figure}

The zero output fields, i.e., $\varepsilon _{outR+}=\varepsilon _{outL+}=0,$
occur under the following conditions
\begin{eqnarray}
\varepsilon _{L}=\varepsilon _{R},\gamma _{1}=\gamma _{2}=2\kappa , \lambda
^{2}=\frac{1}{2}|G|^{2}-\kappa ^{2}.
\end{eqnarray}
Thus there are three channels at
\begin{eqnarray}
D_{0} &=&0,  \notag \\
D_{\pm } &=&\pm \sqrt{|G|^{2}+\lambda ^{2}-3\kappa ^{2}}=\pm \sqrt{\frac{3}{%
2 }|G|^{2}-4\kappa ^{2}},
\end{eqnarray}
for the coherent perfect absorption, implying that the probe lights cannot
be reflected or transmitted from this optomechanical system, but entirely
absorbed. This is due to a perfect destructive interference between the two
probe lights along opposite directions. The energy of the probe lights will
be finally dissipated via the vibrational decay of the NRs and the
thermal-photon decay in the optomechanics, as discussed in detail later. As
a result, this optomechanical system can be used to realize the
multi-channel inverse OMIT (Fig. \ref{fig2}) with the essential prerequisite
of the optomechanical normal-mode splitting \cite{NJP-16-033023}.

For the detuning cases of $D_{\pm }=\pm \sqrt{\frac{3}{2}|G|^{2}-4\kappa^{2}}
$, the effective coupling rate should follow $|G|\geq \sqrt{\frac{8}{3}}
\kappa $. $D_{0}=D_{\pm }=0$ is a special case representing only a single
channel involved in the inverse OMIT when $|G|=\sqrt{\frac{8}{3}}\kappa $
and $\lambda =\frac{1}{\sqrt{3}}\kappa $. Considering the general cases with
$|G|>\sqrt{\frac{8}{3}}\kappa $, there are three channels as presented in
Eq. (13), corresponding to the three injected probe lights at the
frequencies $\omega _{p}=\omega _{c}+\omega _{m}$ and $\omega _{p}=\omega
_{c}+\omega _{m}+D_{\pm }$ with $D_{\pm }=\pm \sqrt{\frac{3}{2}
|G|^{2}-4\kappa ^{2}}$. For example, in the case of $|G|=2\kappa $ and $%
\lambda =\kappa $, the inverse OMIT effect can be observed at $D_{0}=0$ and $%
D_{\pm }=\pm \sqrt{2}\kappa $, corresponding to the three injected probe
lights at the frequencies $\omega _{p}=\omega _{c}+\omega _{m}$ and $\omega
_{p}=\omega _{c}+\omega _{m}\pm \sqrt{2}\kappa $, respectively. Moreover,
these two additional windows become more separate with the increase of both
the effective radiation coupling $|G|$ and the corresponding Coulomb
coupling $\lambda =\sqrt{\frac{1}{2}|G|^{2}-\kappa ^{2}}$, as demonstrated
in Fig. \ref{fig2}.

\section{Discussion}

\subsection{Multi-channel inverse OMIT with two non-identical NRs}

The two identical NRs considered above are theoretically simplified, but
rarely existing experimentally. To release this stringent condition, we
consider below the multi-channel inverse OMIT with non-identical NRs.

\begin{figure}[tbp]
\centering
\includegraphics[width=8cm]{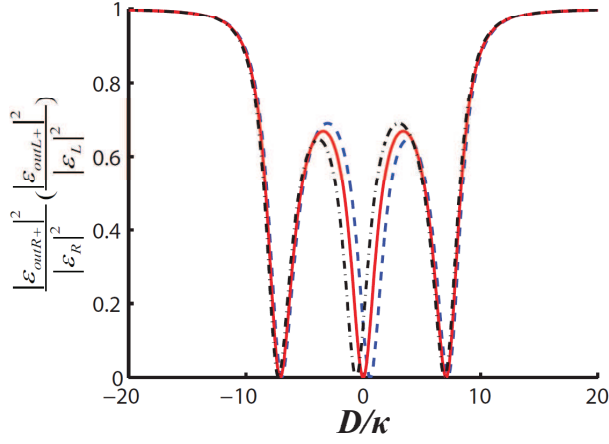}
\caption{The normalized output probe photon number $|\frac{
\protect\varepsilon _{outR+}}{\protect\varepsilon _{R}}|^{2}$ $(|\frac{
\protect\varepsilon _{outL+}}{\protect\varepsilon _{L}}|^{2})$ as a function
of the probe detuning $D/\protect\kappa$, where $D=\protect\omega _{p}-%
\protect\omega _{c}-\protect\omega_{m}$ for identical NRs and $D=\protect%
\omega _{p}-\protect\omega _{c}-\protect\omega_{1}$ for non-identical case.
The red solid, black dashed-dotted, and blue dashed curves are for $\protect%
\omega_{2}= \protect\omega_{1}$, $\protect\omega _{2}=0.8\protect\omega_{1}$
and $\protect\omega_{2}=1.2\protect\omega _{1}$, respectively, if $|G|=6%
\protect\kappa$. }
\label{fig3}
\end{figure}

For two charged NRs with different frequencies, as plotted in Fig. \ref{fig3}
, the multi-channel inverse OMIT occurs with some window shifts with respect
to the case of identical NRs, turning it to be asymmetric for the curves of
the normalized probe photon number versus the probe detuning. It can be
understood from the bright mechanical mode $b=b_{2}\sin \theta +b_{1}\cos
\theta $ and the dark one $d=b_{2}\cos \theta -b_{1}\sin \theta $ with $\tan
\theta =[(\omega _{2}-\omega _{1})+\sqrt{4\lambda ^{2}+(\omega _{2}-\omega
_{1})^{2}}]/(2\lambda )$, which are the diagonalized orthogonal modes of the
two coupled mechanical modes $b_{1}$ and $b_{2}$. These bright and dark
modes can effectively couple to the optical mode with the strengths $|G|\cos
\theta $ and $|G|\sin \theta $, respectively. In contrast to the case of $%
\omega_{1}=\omega_{2}$ with both the bright and dark modes sharing the same
coupling strength $|G|/\sqrt{2}$, the effective couplings for the bright and
dark modes are different in the case of $\omega _{1}\neq \omega _{2}$.
Different from the symmetric curves in the case of $\omega _{1}=\omega_{2}$,
the curves of the normalized probe photon number versus the probe detuning
move leftward if $\omega _{1}>\omega _{2}$ and rightward if $\omega
_{1}<\omega _{2}$ (see Fig. \ref{fig3}). This implies that the middle
channel of this multi-channel inverse OMIT is not always fixed, but variable
if we appropriately tune the frequencies of the NRs, as in \cite%
{JPCM-25-142201}.

More differences can be found in the discussion below from the comparison
between the identical and non-identical NRs.

\begin{table}[tbp]
\caption{The relationship among the normalized output probe photon numbers ($%
\protect\varepsilon _{outR+}$ and $\protect\varepsilon _{outL+}$), the
intracavity probe photon numbers ($\frac{4\protect\kappa ^{2}}{|\protect%
\varepsilon _{L}|^{2}+|\protect\varepsilon _{R}|^{2}}|\protect\delta %
c_{+}|^{2}$), and the mechanical excitations ($\frac{4\protect\kappa ^{2}}{|
\protect\varepsilon _{L}|^{2}+|\protect\varepsilon _{R}|^{2}}|\protect\delta %
b_{1+}|^{2}$, $\frac{4\protect\kappa ^{2}}{|\protect\varepsilon _{L}|^{2}+|
\protect\varepsilon _{R}|^{2}}|\protect\delta b_{2+}|^{2}$ and the
summation) for different effective radiation $|G|$ and Coulomb coupling
strengths $\protect\lambda $ in the inverse OMIT. Part I presents the middle
channel $D=D_{0}$ and part II for the side channels $D=D_{\pm }$. We
consider two non-identical NRs with $\protect\omega _{1}=1.2\protect\kappa $
and $\protect\omega _{2}=\protect\kappa $ and the identical case with $%
\protect\omega _{1}=\protect\omega _{2}=\protect\kappa $. The values in
parentheses are for the identical case.}\centering
\resizebox{\textwidth}{!}{\begin{tabular}{|c|c|c|c|c|c|c|c|c|c|}
\hline
& $D/\kappa $ & $|G|/\kappa $ & $\lambda /\kappa $ & $\varepsilon _{outR+}$ & $
\varepsilon _{outL+}$ & $\frac{4\kappa ^{2}}{|\varepsilon
_{L}|^{2}+|\varepsilon _{R}|^{2}}|\delta c_{+}|^{2}$ & $\frac{4\kappa ^{2}}{
|\varepsilon _{L}|^{2}+|\varepsilon _{R}|^{2}}|\delta b_{1+}|^{2}$ & $\frac{
4\kappa ^{2}}{|\varepsilon _{L}|^{2}+|\varepsilon _{R}|^{2}}|\delta
b_{2+}|^{2}$ & $\frac{4\kappa ^{2}}{|\varepsilon _{L}|^{2}+|\varepsilon
_{R}|^{2}}(|\delta b_{1+}|^{2}+|\delta b_{2+}|^{2})$ \\ \hline
& 0.198 (0) & $2$ & $1$ &  &  & 0.503 (0.5) & 0.4985 (0.5) & 0.4985 (0.5) &
0.997(1.0) \\ \cline{2-4}\cline{7-10}
I & 0.140 (0) & $4$ & $\sqrt{7}$ & 0 (0) & 0 (0) & 0.501 (0.5) & 0.125(0.125)
& 0.874(0.875) & 0.999(1.0) \\ \cline{2-4}\cline{7-10}
& 0.136 (0) & $6$ & $\sqrt{17}$ &  &  & 0.518 (0.5) & 0.006(0.056) &
0.976(0.944) & 0.982(1.0) \\ \hline
& $\begin{array}{c}
+1.417 \\
-1.415\end{array}$($\pm \sqrt{2}$) & $2$ & $1$ &  &  & $\begin{array}{c}
0.502 \\
0.500\end{array}$(0.5) & $\begin{array}{c}
0.711 \\
0.783\end{array}$(0.75) & $\begin{array}{c}
0.287 \\
0.217\end{array}
$(0.25) & $\begin{array}{c}
0.998 \\
1.000\end{array}
$(1.0) \\ \cline{2-4}\cline{7-10}
II & $\begin{array}{c}
+4.629 \\
-4.358\end{array}
$($\pm 2\sqrt{5}$) & $4$ & $\sqrt{7}$ & 0 (0) & 0 (0) & $\begin{array}{c}
0.492 \\
0.508\end{array}
$(0.5) & $\begin{array}{c}
0.742 \\
0.757\end{array}
$(0.75) & $\begin{array}{c}
0.266 \\
0.235\end{array}
$(0.25) & $\begin{array}{c}
1.008 \\
0.992\end{array}
$(1.0) \\ \cline{2-4}\cline{7-10}
& $\begin{array}{c}
+7.227 \\
-6.943\end{array}
$($\pm 5\sqrt{2}$) & $6$ & $\sqrt{17}$ &  &  & $\begin{array}{c}
0.494 \\
0.506\end{array}
$(0.5) & $\begin{array}{c}
0.746 \\
0.754\end{array}
$(0.75) & $\begin{array}{c}
0.261 \\
0.240\end{array}
$(0.25) & $\begin{array}{c}
1.006 \\
0.994\end{array}
$(1.0) \\ \hline
\end{tabular}} 
\end{table}

\subsection{The energy distribution}

We analyze below the paths of the energy dissipation during the inverse OMIT
process. To identify the thermal dissipation in the inverse OMIT, we
calculate the intracavity probe photon number $|\delta c_{+}|^{2}$ and the
quantum excitation of the thermal phonons $|\delta b_{1+}|^{2}$ ($|\delta
b_{2+}|^{2}$) in NR$_1$ (NR$_2$).

Using the fluctuation operators in Eq. (\ref{M2}), we obtain the normalized
intracavity probe photon number
\begin{equation}
\frac{4\kappa ^{2}}{|\varepsilon _{L}|^{2}+|\varepsilon _{R}|^{2}}|\delta {c}
_{+}|^{2}=0.5,  \label{M5}
\end{equation}
which is the ratio of the probe photon number $|\delta c_{+}|^{2}$ versus
the sum of the probe photon numbers $|\frac{\varepsilon _{L}}{2\kappa }
|^{2}+|\frac{\varepsilon _{R}}{2\kappa}|^{2}$ without the coupling field. By
a similar way, the corresponding normalized mechanical excitations of the
charged NR$_{1}$ and NR$_{2}$ for different channels, in units of the sum of
the probe photon numbers, are given, respectively, by
\begin{equation}
\frac{4\kappa ^{2}}{|\varepsilon _{L}|^{2}+|\varepsilon _{R}|^{2}}|\delta {b}
_{1+}|^{2}=
\begin{cases}
\frac{2\kappa ^{2}}{|G|^{2}}, & D_{0}=0 \\
0.75, & D_{\pm }=\pm \sqrt{\frac{3}{2}|G|^{2}-4\kappa ^{2}},%
\end{cases}
\label{M6}
\end{equation}
and
\begin{equation}
\frac{4\kappa ^{2}}{|\varepsilon _{L}|^{2}+|\varepsilon _{R}|^{2}}|\delta {b}
_{2+}|^{2}=
\begin{cases}
1-\frac{2\kappa ^{2}}{|G|^{2}}, & D_{0}=0 \\
0.25, & D_{\pm }=\pm \sqrt{\frac{3}{2}|G|^{2}-4\kappa ^{2}}.%
\end{cases}
\label{M7}
\end{equation}
Equations (\ref{M6}) and (\ref{M7}) present independent thermal dissipations for
the probe lights with different frequencies. Due to this fact, the
multi-channel inverse OMIT can occur simultaneously in the three channels
with different dissipations.
\begin{figure}[tbp]
\centering
\includegraphics[width=10cm]{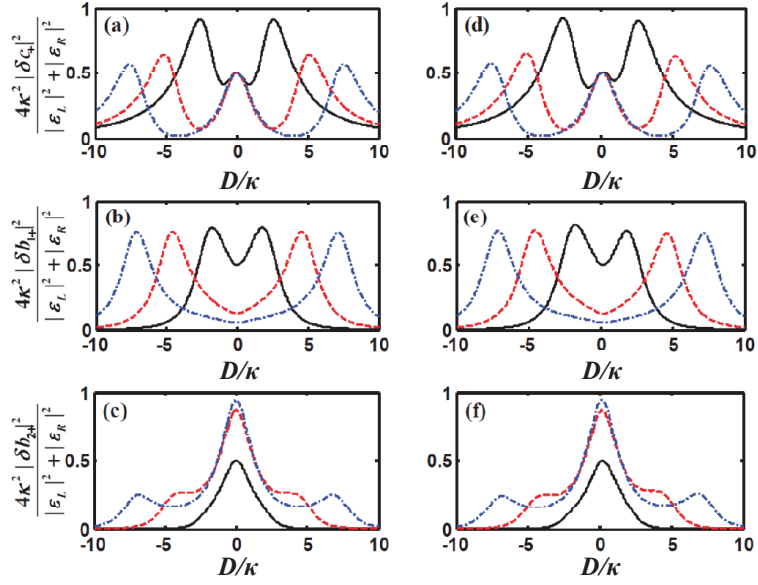}
\caption{Comparison of identical and non-identical NRs in the
variation of photon and phonon numbers with respect to the probe detuning $%
D/ \protect\kappa$. The panels: (a) and (d) for the normalized probe photon
number $\frac{4\protect\kappa ^{2}}{|\protect\varepsilon _{L}|^{2}+|\protect%
\varepsilon _{R}|^{2}}|\protect\delta {c}_{+}|^{2}$; (b) and (e) for the
normalized mechanical excitation $\frac{4\protect\kappa ^{2}}{|\protect%
\varepsilon _{L}|^{2}+|\protect\varepsilon _{R}|^{2}}|\protect\delta {b}
_{1+}|^{2}$; (c) and (d) for $\frac{ 4\protect\kappa ^{2}}{|\protect%
\varepsilon _{L}|^{2}+|\protect\varepsilon_{R}|^{2}}|\protect\delta {b}
_{2+}|^{2}$. (a), (b) and (c) plot identical NRs with $D=\protect\omega_{p}-
\protect\omega_{c}-\protect\omega_{m}$, while (d), (e) and (f) present
non-identical NRs ($\protect\omega_{2}=1.2\protect\omega_{1}$) with $D=
\protect\omega _{p}-\protect\omega_{c}-\protect\omega_{1}$. The black solid,
red dashed, and blue dashed-dotted curves are for pumping rates $|G|=2%
\protect\kappa$, $4\protect\kappa$, $6\protect\kappa$, respectively.}
\label{fig4}
\end{figure}

From Figs. \ref{fig4} and Table 1, we find in the case of identical NRs
that, when the inverse OMIT takes place, the sum of the mechanical
excitations [$\frac{4\kappa^{2}}{|\varepsilon_{L}|^{2}+|\varepsilon
_{R}|^{2} }(|\delta b_{1+}|^{2}+|\delta b_{2+}|^{2})\equiv 1$] is always
double of the intracavity probe photon number [$\frac{4\kappa ^{2}}{%
|\varepsilon _{L}|^{2}+|\varepsilon _{R}|^{2}}|\delta c_{+}|^{2}=0.5$]. This
implies that the energy distribution in the two NRs and the cavity field
always remains with the ratio $2:1$. Besides, with increase of $|G|$ and $%
\lambda$, the phonon distribution in the two NRs varies in different ways
conditional on the channels. For two different NRs, there is a small
deviation with respect to the identical case, and only the middle channel
satisfies the condition for the inverse OMIT. These characteristics of our
multi-channel inverse OMIT are very different from in previous schemes \cite%
{NJP-16-033023, OE-22-4886}.

\subsection{Measurement of the relative phase in a unilateral inverse OMIT}

Since the inverse OMIT is resulted from the perfect destructive interference
between two probe lights along opposite directions \cite{NJP-16-033023}, any
imperfection, such as a phase difference between the two probe lights, would
lead to deviation from the perfect destructive interference. As such, it
would be interesting to explore the possibility of detecting the difference
between the two probe lights in an imperfect inverse OMIT.

\begin{figure}[tbp]
\centering\includegraphics[width=10cm]{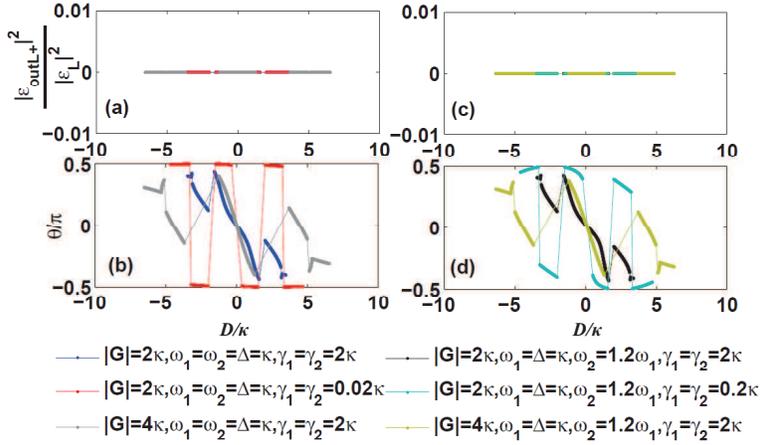}
\caption{(a) and (c) The strength of the output light from
the left-hand side of the optomechanical cavity with different probe light
detunings, which remains unchanged for different parameter values; (b) and
(d) The phase $\protect\theta$ of the unilateral inverse OMIT for different
probe light detunings. The left-hand side panels are for identical NRs ($%
\protect\omega_{2}=\protect\omega_{1}$) with $D=\protect\omega _{p}-\protect%
\omega _{c}-\protect\omega_{m}$, while the right-hand side panels for
non-identical NRs ($\protect\omega_{2}=1.2\protect\omega_{1}$) with $D=
\protect\omega _{p}-\protect\omega _{c}-\protect\omega_{1}$.}
\label{measurement}
\end{figure}

After a relative phase $\theta$ is introduced in the probe light input from
the right-hand side of the cavity, the inverse OMIT observed in the
left-hand side takes the form
\begin{equation}
\begin{array}{ccc}
\varepsilon _{outL+}=\frac{2\kappa (\varepsilon _{L}+\varepsilon
_{R}e^{i\theta })}{2\kappa +i(\Delta -\delta )+\frac{|G|^{2}}{\frac{{\gamma
_{1}}}{2}-i(\delta -\omega _{1})+\frac{{\lambda ^{2}}}{{\frac{{\gamma _{2}}}{
2}}-i(\delta -{\omega _{2}})}}}-\varepsilon _{L} & = & 0.%
\end{array}
\label{M54}
\end{equation}
In contrast to the same output lights ($\varepsilon _{outL+}=\varepsilon
_{outR+}$) from both sides of the cavity, the inverse OMIT involving a
relative phase outputs the lights with $\varepsilon_{outL+}\neq
\varepsilon_{outR+}$, implying that the inverse OMIT, if occurring, is
observed only on one side of the cavity (i.e., an unilateral OMIT). In such
an unilateral inverse OMIT, the relative phase $\theta$ is found to be
monotonously varying with the detuning $D$ within some parameter regimes,
which can be employed for evaluating $\theta$ (see Fig.\ref{measurement}).

Provided that the strengths of the two probe lights are the same ($%
\varepsilon _{L}=\varepsilon _{R}$), the above equation is reduced to
\begin{equation}
\begin{array}{ccc}
\frac{|G|^{2}}{\frac{{\gamma _{1}}}{2}-i(\delta -\omega _{1})+\frac{{\lambda
^{2}}}{{\frac{{\gamma _{2}}}{2}}-i(\delta -{\omega _{2}})}}-i(\delta -\Delta
) & = & 2\kappa e^{i\theta }.%
\end{array}
\label{M64}
\end{equation}
Straightforward deduction using the relations among $D$, $\delta $ and $%
\Delta $ yields that, the relative phase $\theta $ is a function of the
detuning $D$, as plotted in Fig. \ref{measurement}(b) and Fig. \ref{measurement}(d), and not all the
frequencies of the probe lights can generate the unilateral inverse OMIT
effect.

For a precision measurement of $\theta $, choosing qualified regimes of the
parameters, e.g., $D/\kappa \in \lbrack -1.5,1.5]$, is necessary to obtain a
monotonous change of $\theta $ with $D$. Besides, for the measurement to be
more precise, we expect a large change of $D$ for tiny variation of $\theta$%
, implying a small slope of $\Delta \theta /\Delta D$. As such, smaller
radiation coupling is optimal [see the curves in Fig. \ref{measurement}(b)
with $|G|=2\kappa ,4\kappa $ and note the lower limit $|G|\geq \sqrt{8/3}
\kappa $]. In comparison with the identical NRs
[in Fig. \ref{measurement}(b)], the curves for the non-identical NRs [in Fig. \ref{measurement}(d)]
own smaller slopes, implying a better measurement. For example, in the case
of $D/\kappa \in \lbrack -0.01,0.01]$, the measurement sensitivity $\Delta
D/\Delta \theta $ is 6.3 MHz/$\mathrm{rad}$ for the blue curve in Fig. \ref
{measurement}(b), and 7.7 MHz/$\mathrm{rad}$ for the black curve in Fig. \ref
{measurement}(d). So elaborately choosing different NRs can be favorable for
enhancing the measurement precision of $\theta $. By numerical simulation of
Eq. (18), we find the largest measurement sensitivity $\Delta D/\Delta
\theta =$8.3 MHz/$\mathrm{rad}$ at $\omega _{2}/\omega _{1}=1.346$, since
the unilateral inverse OMIT would disappear once $\omega _{2}/\omega
_{1}>1.34$.

\begin{figure}[tbp]
\centering\includegraphics[width=8cm]{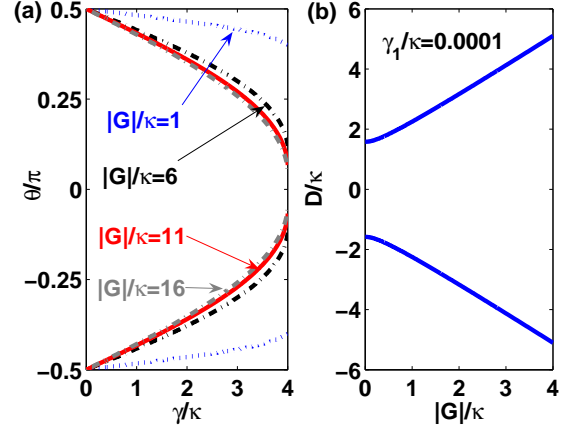}
\caption{(a) The decay of the NR versus the relative phase
between two probe lights for different effective radiation couplings; (b)
The detuning of the probe field from the cavity resonance frequency verses
the effective radiation coupling, where $\Delta =\protect\omega _{1}$ and $D=%
\protect\delta -\protect\omega _{1}=\protect\delta -\Delta $.}
\label{oneMR}
\end{figure}

Moreover, this unilateral inverse OMIT can also be applied to typical
optomechanical cavity with only one NR. For example, when the Coulomb
coupling is decoupled, Eq. (\ref{M64}) can be reduced to
\begin{equation}
\begin{array}{ccc}
\frac{|G|^{2}}{\frac{{\gamma _{1}}}{2}-iD}-iD & = & 2\kappa e^{i\theta },
\end{array}
\label{M65}
\end{equation}
for $\Delta =\omega _{1}$ and $D=\delta -\omega _{1}=\delta -\Delta $. Then
we obtain the corresponding detunings for the unilateral inverse OMIT as
\begin{equation}
D_{\pm }=\pm \sqrt{\frac{1}{8}(8|G|^{2}+(16\kappa ^{2}-{\gamma _{1}^{2})}+
\sqrt{16|G|^{2}(16\kappa ^{2}-{\gamma _{1}^{2}})+(16\kappa ^{2}+{\gamma
_{1}^{2}})}}.  \label{M66}
\end{equation}
With the assistance of Eq. (\ref{M65}), the decay of the NR versus the
relative phase between two probe lights is calculated with respect to
different effective radiation couplings, as plotted in Fig. \ref{oneMR}(a).
It implies that the unilateral inverse OMIT can be observed for any decay
rate of the NR, which is a great improvement on the inverse OMIT compared to
that in \cite{NJP-16-033023}, where the inverse OMIT can only be
achieved when $4\kappa ={\gamma _{1}}$. Besides, we numerically treat
Eq.(\ref{M66}) in Fig. \ref{oneMR}(b), showing that the detuned frequency for
the unilateral inverse OMIT is almost linear to the effective radiation
coupling with a ratio $\Delta |G|/\Delta D=1.016$, and thus the effective
radiation coupling can be identified by the detuned frequency in this way.

\section{The experimental feasibility}

We exemplify the unilateral inverse OMIT for a brief discussion about the
experimental feasibility of our scheme. In terms of the experimental
parameters in \cite{Nature-460-724}, we set following parameter values available
using current techniques: The frequencies of the NRs $\omega_{1}\simeq\omega_{2}=2\pi\times 947$\ kHz,
$m_{1}\simeq m_{2}=145$ ng, the cavity wavelength $\lambda_{c}\equiv 2\pi c/\omega
_{c}=1064$ nm, the cavity length $L=25$\ mm, and the cavity decay
rate $\kappa =2\pi\times 215$ kHz. Then the effective radiation coupling is $
|G|=g_{0}|c_{s}|=\hbar\omega_{c}/L|c_{s}|\geq 2\pi\times 351$ kHz
corresponding to the Coulomb coupling strength $\lambda \geq 2\pi\times 124$ kHz
\cite{pra-86-053806,pra-90-043825,pra-91-063827,np-6-707,science-304-74}
and the strength of the driven light field $\varepsilon _{c}=\sqrt{
2\wp _{c}\kappa /\hbar \omega _{c}}$\ with a input power $\wp _{c}\geq 0.037$
mW.

The prerequisite of observing an inverse OMIT is a near red-sideband resonance ($\Delta\simeq\omega_{1}$).
Considering the requirement for sideband resolution, we may rewrite the condition of the near red-sideband resonance more specifically
as $0\leq |\Delta -\omega _{1}|\ll \omega _{1}$. In the case of an exactly red-sideband resonance, we have a symmetrical inverse OMIT.
When there is a tiny deviation from the exactly red-sideband resonance, but satisfying $0<|\Delta -\omega_{1}|\ll\omega_{1}$, we
have an asymmetrical profile and in this case the inverse OMIT still occurs. The possibility lies in following points:
i) The OMIT originates from the quantum interference generally with an asymmetrical profile, and the symmetrical profile
is a special case; ii) The phase shift in the destructive interference is generated by the ratio between the dispersion
and the absorption caused by the quantum interference. As such, our scheme can work
within the regime: $0<|\Delta -\omega_{1}|\ll\omega_{1}$.

\section{Conclusion}

In summary, we have investigated a tunable multi-channel inverse OMIT in the
optomechanical system with the assistance of the Coulomb interaction between
two charged NRs. Our results have shown both analytically and numerically
three channels for perfectly absorbing the input probe fields at different
frequencies in such a system, which makes it possible to select a desired
frequency of inverse OMIT by adjusting effective radiation coupling rate and
the corresponding Coulomb coupling strength. Some applications have been
discussed based on the considered model. We believe that our study would be
useful for further understanding the inverse OMIT and exploring the
applications of the inverse OMIT.

Based on the schematic in Fig. \ref{fig1}, our scheme can be extended to other systems such as a
waveguide optomechanics \cite{nature-472-69} and the Coulomb coupling
between two NRs by a bias voltage gate \cite{pra-72-041405}. We have noted the opto-mechanical experiments reported
recently with NRs coupled by tunable optical coupling \cite{nphotonics} and
fixed elastic coupling \cite{apl-105-014108}. Replacing the Coulomb coupling
by those couplings, our model can immediately apply to those opto-mechanical
systems in \cite{nphotonics,apl-105-014108}. In addition, we are also aware
of a recent work for a multi-channel inverse OMIT by confining many NRs in a
single cavity \cite{arxiv}. The idea is very interesting, but much more
difficult to achieve experimentally than our scheme.

\section*{Acknowledgments}

QW thanks Lei-Lei Yan, Yin Xiao and Peng-Cheng Ma for their helps in
numerical simulation. JQZ thanks for the help from Prof. Zheng-Yuan Xue.
This work is supported by the National Natural Science
Foundation of China (Grants Nos. 91121023, 61378012, 60978009, 11274352,
91421111 and 11304366), the SRFDPHEC (Grant No. 20124407110009), National
Fundamental Research Program of China (Grants Nos. 2011CBA00200,
2012CB922102 and 2013CB921804), the PCSIRT (Grant No. IRT1243), and China
Postdoctoral Science Foundation (Grants Nos. 2013M531771 and 2014T70760).

\end{document}